\documentclass[a4paper,11pt]{article}
\pdfoutput=1 

\usepackage{jheppub,xcolor} 

\usepackage[T1]{fontenc} 

\renewcommand{\a}{\alpha}
\renewcommand{\b}{\beta}
\renewcommand{\d}{\delta}
\newcommand{\g}{\gamma}
\newcommand{\p}{\partial}

\newcommand{\be}{\begin{equation}}
\newcommand{\ee}{\end{equation}}
\newcommand{\bea}{\begin{align}}
\newcommand{\eea}{\end{align}}

\title{\rm \bf \Huge Supertrace formulae for nonlinearly realized supersymmetry}

\author{Divyanshu Murli}
\author{and Yusuke Yamada}
\affiliation{Stanford Institute for Theoretical Physics and Department of Physics, Stanford University, Stanford, CA 94305, USA}

\abstract{We derive the general supertrace formula for a system with $N$ chiral superfields and one nilpotent chiral superfield in global and local supersymmetry. The nilpotent multiplet is realized by taking the scalar-decoupling limit of a chiral superfield breaking supersymmetry spontaneously. As we show, however, the modified formula is not simply related to the scalar-decoupling limit of the supertrace in linearly-realized supersymmetry. We also show that the supertrace formula reduces to that of a linearly realized supersymmetric theory with a decoupled sGoldstino if the Goldstino is the fermion in the nilpotent multiplet.}

\begin{document} 
\maketitle
\flushbottom

\section{Introduction}
Spontaneous breaking of supersymmetry is one of the most important issues for realistic model building in supersymmetric theories. The nonlinear realization is a useful tool for the description of the physics below the energy scale of spontaneous SUSY breaking, which manifests in the form of constrained superfields. Appropriate constraints reduce the physical degrees of freedom in a given superfield in a supersymmetric manner; this corresponds to the decoupling limit of heavy particles which acquire their mass from supersymmetry breaking effects~\cite{Komargodski:2009rz,DallAgata:2015zxp,Ferrara:2016een,Kallosh:2016hcm,DallAgata:2016syy,Cribiori:2017ngp}. 

A nilpotent chiral superfield $\hat S(x,\theta)$ satisfying $\hat{S}^2=0$ is particularly important to describe supersymmetry breaking~\cite{Volkov:1972jx,Volkov:1973ix,Rocek:1978nb,Ivanov:1978mx,Lindstrom:1979kq,Casalbuoni:1988xh,Komargodski:2009rz}. The nilpotent condition has nontrivial solution if and only if $F^S\neq0$, which means supersymmetry is broken by $\hat{S}$. Interestingly, such a nilpotent superfield appears in the low energy effective action of an anti-D3 brane in some classes of superstring models~\cite{McGuirk:2012sb,Kallosh:2014wsa,Bergshoeff:2015jxa,Kallosh:2015nia,Vercnocke:2016fbt,Kallosh:2016aep}. Such an anti-D3 brane effect plays an important in realizing de Sitter vacua in string theory~\cite{Kachru:2003aw}. The application of the nilpotent superfield to inflationary cosmology has also been studied~\cite{Antoniadis:2014oya,Ferrara:2014kva,Kallosh:2014via,DallAgata:2014qsj,Kallosh:2014hxa}. The absence of an independent scalar $S$ in the $\hat{S}$ superfield has an advantage in such model building.

From this perspective, it would be important to understand the properties of nonlinear supersymmetry. In linearly realized supersymmetry, the supertrace mass formula~\cite{Ferrara:1979wa,Cremmer:1982en,Grisaru:1982sr} shows one of the most interesting properties; it takes a simple form described in terms of the underlying K\"ahler geometry. This formula also has practical uses, e.g. loop corrections to the vacuum energy. Recently, the supertrace formula in curved spacetime was investigated in \cite{Ferrara:2016ntj}.

In this work, we will derive the supertrace mass formula for a system with a nilpotent superfield in global and local supersymmetry models. The complete component action for such a system was recently shown in \cite{Bergshoeff:2015tra,Hasegawa:2015bza,Kallosh:2015sea,Kallosh:2015tea,Schillo:2015ssx}. It turns out that the presence of a nilpotent superfield leads not only to the absence of the scalar $S$ but also additional terms in the Lagrangian, which seem to break the geometric form of the standard supergravity action. For these reasons, one expects that the supertrace mass formula is also different from the standard one. Indeed, as we will show, the formula consists of a standard geometric part and a non-geometric part originating from the nilpotent superfield.

The remaining part of this paper is organized as follows. In Sec.~\ref{sec2}, we briefly review the general features of a system with a nilpotent superfield. We then derive the general form of the modified supertrace mass formula, which will be used in the following sections. We construct the explicit modified supertrace mass formula for global supersymmetry in Sec.~\ref{sec3}, and for supergravity in Sec.~\ref{sec4}. Finally we conclude in Sec.~\ref{sec5}.
\section{Nilpotent condition and mass correction}\label{sec2}
In this section we briefly review the properties of the nilpotent superfield and discuss possible corrections to the mass formulae for both bosons and fermions. The nilpotent condition $\hat S^2(x,\theta)=0$ on a chiral superfield,
\begin{equation}
\hat S(x,\theta)= S(x) + \sqrt{2}\theta \chi^S(x) + \theta^2 F^S(x),
\end{equation} leads to~\cite{Rocek:1978nb,Ivanov:1978mx,Lindstrom:1979kq,Casalbuoni:1988xh,Komargodski:2009rz}
\begin{equation}
S=\frac{\chi^{S}\chi^S}{2F^S},
\end{equation}
where $S, \ \chi^{S}$ and $F^{S}$ are the scalar, a spinor and an auxiliary scalar field components of $S(x,\theta)$, respectively. This equation shows the following properties of the nilpotent superfield: 1. the scalar $S$ is not a physical degree of freedom, and 2. a linear term of $S$ in Lagrangian behaves like a mass term of $\chi^S$.

From the first observation, we find that the scalar mass formula should be modified as follows. Since $S$ is not a dynamical degree of freedom and should be projected out as $S=0$, the sum of scalar masses is given by
\begin{align}
\left(\frac12\text{tr} M_0^2\right)_{\rm nl}= &g^{a\bar b} V_{a\bar b}\biggl|_{S=0}\nonumber\\
=&\left[g^{\alpha\bar \beta}V_{\alpha\bar \beta}-g^{S\bar b}V_{S\bar b}-g^{a\bar S}V_{a\bar S}-g^{S\bar S}V_{S\bar S}\right]\biggl|_{S=0}\label{gsmf}
\end{align}
where $V$ is the scalar potential; the roman indices $a,b$ denote all scalar fields aside from $S$, the greek indices $\alpha,\beta$ run over all scalar fields including $S$ and subscripts on $V$ denote differentiation with respect to scalar fields. $g^{\alpha \bar \beta}$ denotes the inverse of K\"ahler metric. Here, $V$ and its derivatives should be calculated assuming $S \neq 0$; in the end we take the projection $S=0$ represented by $|_{S=0}$. The first term of this formula $g^{\alpha\bar \beta}V_{\alpha\bar \beta}\equiv(\text{tr}M_{0}^2)_{\text{lin}}$ corresponds to the total scalar mass formula for the linearly realized supersymmetric case, where $\hat S$ is unconstrained.

We also find the corrections to the fermion mass formula from the second observation made above. Let us formally write a fermion mass in linear supersymmetry as $m_{\alpha\beta}$, which will be defined in the following sections. Since the linear term of $S$ in $V$ becomes the fermion mass term of $\chi^{S}$, the fermion mass formula is corrected as
\begin{equation}
m'_{\alpha\beta}=m_{\alpha\beta}+\delta^{S}_{\alpha}\delta^{S}_{\beta}\frac{V_S}{F^S}\biggl|_{S=0},\label{deffm}
\end{equation}
where we have defined the mass term of the fermion in Lagrangian as $-\frac{1}{2}m'_{\alpha\beta}\chi^{\alpha}\chi^{\beta}$. Thus, the fermion mass trace is given by
\begin{align}
\left(\text{tr} M_{1/2}^{2}\right)_{\rm nl}=&g^{\alpha\bar\beta}g^{\gamma\bar\delta}m'_{\alpha\gamma}\bar{m}'_{\bar\beta\bar\delta}\biggl|_{S=0}\nonumber\\
=&\left[g^{\alpha\bar\beta}g^{\gamma\bar\delta}m_{\alpha\gamma}\bar{m}_{\bar\beta\bar\delta}+g^{S\bar \alpha}g^{S\bar \beta}\bar{m}_{\bar \alpha\bar \beta}\frac{V_S}{F^S}+g^{a\bar S}g^{b\bar S}m_{ab}\frac{V_{\bar S}}{\bar{F}^S}+(g^{S\bar S})^2\left|\frac{V_S}{F^S}\right|^2\right]\biggl|_{S=0}\nonumber\\
=&\left[(\text{tr}M_{1/2}^2)_{\text{lin}}+\Delta M_{1/2}^{2}\right]\biggl|_{S=0},
\end{align}
where $\left(\text{tr}M_{1/2}^2\right)_{\text{lin}}\equiv g^{\alpha\bar\beta}g^{\gamma\bar\delta}m_{\alpha\gamma}\bar{m}_{\bar\beta\bar\delta}$ and 
\begin{equation}
\Delta M_{1/2}^{2}\equiv g^{S\bar a}g^{S\bar b}\bar{m}_{\bar a\bar b}\frac{V_S}{F^S}+g^{a\bar S}g^{b\bar S}m_{ab}\frac{V_{\bar S}}{\bar{F}^S}+(g^{S\bar S})^2\left|\frac{V_S}{F^S}\right|^2.
\end{equation}
The first term $\left(\text{tr} M_{1/2}^{2}\right)_{\text{lin}}$ corresponds to the sum of fermion masses in linearly realized supersymmetry, whereas the second contribution $\Delta M_{1/2}^{2}$ comes from the extra mass term of $\chi^{S}$ induced by the nilpotent condition.

As discussed above, in both the scalar and fermion mass traces, there are corrections originating from the nilpotent superfield $\hat S$. Note that in supergravity there is no correction to the gravitino mass aside from imposing the condition $S=0$. Taking the corrections into account, we find that the relationship between the supertrace mass formulae in linear and nonlinear supersymmetry is given by
\begin{align}
\frac{1}{2}\left(\text{Str}\mathcal{M}^2\right)_{\text{nl}}= &\left(\frac{1}{2} \text{tr}  M_0^2 \right)_{\rm nl}- \left(\text{tr}M_{1/2}^2\right)_{\rm nl}-2m_{3/2}^2\biggl|_{S=0}\nonumber\\
= &\left[\left(\frac{1}{2} \text{tr}M_0^2\right)_{\rm lin} - \left(\text{tr} M_{1/2}^2\right)_{\rm lin}-2m_{3/2}^2 -\frac12\Delta M_0^2-\Delta M_{1/2}^2\right]\biggl|_{S=0}\nonumber\\
= &\frac12\left(\text{Str}\mathcal{M}^2\right)_{\text{lin}}\biggl|_{S=0} -\left(\frac12\Delta M_0^2+\Delta M_{1/2}^2\right)\biggl|_{S=0},\label{gen}
\end{align}
where $\left(\text{Str}\mathcal{M}^2\right)_{\text{lin (nl)}}$ denotes the supertrace mass formula for (non-)linear supersymmetry and 
\begin{equation}
\frac12\Delta M_0^2\equiv g^{S\bar b}V_{S\bar b}+g^{a\bar S}V_{a\bar S}+g^{S\bar S}V_{S\bar S}.
\end{equation}
Note that the spin 3/2 part is absent in global supersymmetry. Using this general relation~\eqref{gen}, we derive the explicit form of the supertrace formulae in global and local supersymmetry in the following section. 

One may see the formula \eqref{gen} as
\begin{equation}
\frac{1}{2}\left(\text{Str}\mathcal{M}^2\right)_{\text{nl}}=\frac12\left(\text{Str}\mathcal{M}^2\right)_{\text{S-decouple}} -\Delta M_{1/2}^2\biggl|_{S=0}.
\end{equation}
Here we have defined
\begin{equation}
\frac12\left(\text{Str}\mathcal{M}^2\right)_{\text{S-decouple}}=\frac12\left(\text{Str}\mathcal{M}^2\right)_{\text{lin}}\biggl|_{S=0} -\frac12\Delta M_0^2\biggl|_{S=0}.
\end{equation}
One might expect that $\frac12\left(\text{Str}\mathcal{M}^2\right)_{\text{S-decouple}}$ is the supertrace formula in a system with a nilpotent superfield, since such a system would correspond to the decoupling limit of the scalar $S$. However, the presence of $\Delta M_{1/2}^2$ shows that it is not the case: the supertrace mass formula is unexpectedly deformed in a nonlinearly-realized-supersymmetric system.
\section{Global supersymmetry}\label{sec3}
In this section, we derive the supertrace mass formula with a nilpotent superfield in global supersymmetry. We consider a system with $N$ chiral superfields ($\hat z^a$) and one nilpotent superfield $\hat S$. Since $\hat{S}$ satisfies the nilpotent condition, the general form of the K\"ahler and super-potential are restricted as
\begin{align}
K=&K_0+K_1S+K_{\bar1}\bar S+K_2S\bar S,\\
W=&W_0+W_1S,
\end{align}
where $K_{0,1,\bar 1,2}$ are functions of scalar fields $z^a$ and $\bar z^{\bar b}$, and $W_{0,1}$ are holomorphic functions of $z^a$. In a linearly realized supersymmetry case, the scalar potential is given by 
\begin{equation}
V = g^{\a \bar{\b} } W_{\a} \bar{W}_{\bar{\b}}.
\end{equation} 
We have to project $S$ out from the actual scalar potential if $S$ is a nilpotent superfield, and then the scalar potential should be
\begin{equation}
\left(V\right)_{\rm nl}=g^{\a \bar{\b} } W_{\a} \bar{W}_{\bar{\b}}\biggl|_{S=0}.
\end{equation}
We have an algebraic relation
\begin{align}\label{Vs}
V_\a &= W_{\a\b} g^{\b\bar\g} \bar{W}_{\bar\g} + \p_\a g^{\b\bar\g}W_{\b} \bar{W}_{\bar\g} \nonumber \\
&=W_{\a\b} (g^{\b \bar{\g}} \bar{W}_{\bar\g}) -  \Gamma^{\d}_{\a\b} W_{\d} (g^{\b \bar{\g}} \bar{W}_{\bar\g}) \nonumber \\
&=   \nabla_\a W_{\b}(g^{\b \bar{\g}} \bar{W}_{\bar\g})   \nonumber \\
& = -m_{\a\b} F^{\b},
\end{align}
where $m_{\a\b}$ is a fermion mass given by
\begin{equation}
m_{\a\b}\equiv \nabla_\a W_\b=W_{\a\b}-\Gamma_{\a\b}^\g W_\g,
\end{equation}
and $\Gamma_{\a\b}^\g=g^{\g\bar\d}g_{\a\b\bar\d}$ is the  K\"ahler connection. Note that, since $S$ satisfies the nilpotent condition, we find $W_{SS}=\Gamma_{SS}^{\a}=0$, and therefore \begin{equation}
m_{SS}\equiv 0.
\end{equation}
A similar calculation yields
\begin{equation}\label{Vss}
V_{\a \bar{\a}} = m_{\a \b} g^{\b \bar{\b}} \bar{m}_{\bar{\a} \bar{\b}} + R_{\a \bar{\a}\b \bar{\b}}F^{\b} \bar{F}^{\bar{\b}},
\end{equation}
where $R_{\a \bar{\a}\b \bar{\b}}=K_{\a\b\bar\a\bar\b}-g^{\g\bar\g}K_{\a\b\g}K_{\bar\a\bar\b\bar\g}$.  Note that these identities also hold under the projection $S=0$.

We derive each term in \eqref{gen} using these identities. The first term of \eqref{gen} corresponds to the usual supertrace mass formula. Using the identity \eqref{Vss}, we find
\begin{align}
\frac12\left(\text{Str}\mathcal{M}^2\right)_{\text{lin}}\biggl|_{S=0}=&\left[g^{\a \bar{\a}} V_{\a \bar{\a} }-m_{\a\b}g^{\a\bar\a}g^{\b\bar\b}\bar m_{\bar\a\bar\b}\right]\biggl|_{S=0}\nonumber\\
=&g^{\alpha\bar\alpha}R_{\a \bar{\a}\b\bar\b} F^{\b} \bar{F}^{\bar{\b}}\biggl|_{S=0}.
\end{align}
The second term of \eqref{gen} can be expressed as
\begin{align}
\frac12\Delta M_0^2=&g^{S\bar a}(m_{Sa}g^{a\bar\a}\bar m_{\bar a\bar\a}+R_{S\bar aa\bar\a} F^a\bar F^{\bar\a})+g^{a\bar S}(m_{a\a}g^{\a\bar a}\bar m_{\bar S\bar a}+R_{a\bar S\a\bar a}F^\a\bar F^{\bar a})\nonumber\\
&+g^{S\bar S}(m_{Sa} g^{a\bar a}m_{\bar S\bar a}+R_{S\bar Sa\bar a}F^a\bar F^{\bar a}),
\end{align}
where we have taken into account $m_{SS}=0$ and $R_{S\bar\a S\bar\b}=0$, which are consequences of the nilpotent condition.
Using the identity~\eqref{Vs}, we find
\begin{equation}
\frac{V_S}{F^S}=-m_{SS}-\frac{F^a}{F^S}m_{Sa}=-\frac{F^a}{F^S}m_{Sa}\label{vsfs}.
 \end{equation}
Note that for a nonlinear superfield $V_S=0$ is not realized dynamically, because of the absence of a dynamical scalar $S$ and hence this quantity does not vanish at the vacuum. 
The third term of \eqref{gen} can then be expressed as
\begin{align}
\Delta M_{1/2}^2=&-g^{S\bar a}g^{S\bar b}m_{\bar a\bar b}\frac{F^a}{F^S}m_{Sa}-g^{a\bar S}g^{b\bar S}m_{ab}\frac{\bar F^{\bar a}}{\bar F^{\bar S}}m_{\bar S \bar a}+(g^{S\bar S})^2\left|\frac{F^a}{F^S}m_{Sa}\right|^2.
\end{align}
Finally, we find the supertrace formula with $N$-chiral superfields and a nilpotent chiral superfield $\hat S$ is
\begin{align}
\frac12\left(\text{Str}\mathcal{M}^2\right)_{\text{nl}}=
&g^{a\bar a}R_{a \bar{a}\a\bar\a} F^{\a} \bar{F}^{\bar{\a}}-g^{S\bar a}m_{Sa}g^{a\bar\a}\bar m_{\bar a\bar\a}-g^{a\bar S}m_{a\a}g^{\a\bar a}\bar m_{\bar S\bar a}-g^{S\bar S}m_{Sa} g^{a\bar a}m_{\bar S\bar a}\nonumber\\
&+g^{S\bar a}g^{S\bar b}m_{\bar a\bar b}\frac{F^a}{F^S}m_{Sa}+g^{a\bar S}g^{b\bar S}m_{ab}\frac{\bar F^{\bar a}}{\bar F^{\bar S}}m_{\bar S \bar a}-(g^{S\bar S})^2\left|\frac{F^a}{F^S}m_{Sa}\right|^2.\label{glgen}
\end{align}
Although combining all terms partially simplifies the total expression, the expression for the supertrace mass formula is still complicated. This is a significant difference between the formula in linear and nonlinear supersymmetry. 

\subsection{Simplification: $F^a=0$}
Let us consider a condition which reduces the complexity of the supertrace formula. The particular difference between the linear and nonlinear supersymmetry comes from the additional mass contribution $\Delta M_{1/2}^2$. Indeed, the relation between $\left(\text{Str}\mathcal{M}^2\right)_{\text{nl}}$ and $\left(\text{Str}\mathcal{M}^2\right)_{\text{S-decouple}}$ becomes
\begin{equation}
\left(\text{Str}\mathcal{M}^2\right)_{\text{nl}}=\left(\text{Str}\mathcal{M}^2\right)_{\text{S-decouple}}+g^{S\bar a}g^{S\bar b}m_{\bar a\bar b}\frac{F^a}{F^S}m_{Sa}+g^{a\bar S}g^{b\bar S}m_{ab}\frac{\bar F^{\bar a}}{\bar F^{\bar S}}m_{\bar S \bar a}-(g^{S\bar S})^2\left|\frac{F^a}{F^S}m_{Sa}\right|^2.
\end{equation}
From this expression, one finds that for $F^a=0$,
\begin{align}
\left(\text{Str}\mathcal{M}^2\right)_{\text{nl}}&=\left(\text{Str}\mathcal{M}^2\right)_{\text{S-decouple}}\nonumber\\
&=\left[g^{a\bar a}R_{a \bar{a}S\bar S} F^{S} \bar{F}^{\bar{S}}-g^{S\bar a}m_{Sa}g^{a\bar\a}\bar m_{\bar a\bar\a}-g^{a\bar S}m_{a\a}g^{\a\bar a}\bar m_{\bar S\bar a}-g^{S\bar S}m_{Sa} g^{a\bar a}m_{\bar S\bar a}\right]\biggl|_{S=0}.
\end{align}
In this case, the supertrace mass formula corresponds to that in linearly realized supersymmetry with decoupling of a scalar $S$. The condition $F^a=0$ means that supersymmetry breaking is caused only by a single superfield $\hat{S}$. In particular for $g_{S\bar{a}}=0$, $\chi^S$ becomes the Goldstino.~\footnote{For a case only with $F^a=0$, $W_a\neq0$ is possible in general. Then, the Goldstino $G$ is a linear combination of the fermions, $G=W_S\chi^{S}+W_a\chi^a$.}

This result is consistent with the observation in \cite{Antoniadis:2011xi}: in the linearly realized supersymmetry models where some superfields have non-vanishing F-terms, the infrared limit of the model does not realize a nilpotent superfield but a constrained superfield $X$ with a cubic nilpotent constraint $X^3=0$.\footnote{Conditions for $X^2=0$ to be valid even in the presence of additional SUSY breaking fields are studied in~\cite{Antoniadis:2012ck}.}

We can further simplify the formula by imposing the vacuum condition $V_a=0$. This vacuum condition leads to
\begin{equation}
V_a=-m_{a\a}F^{\a}=0.
\end{equation}
Since we have assumed $F^{a}=0$ (and $F^S\neq0$), the vacuum condition is equivalent to
\begin{equation}
m_{aS}=0.
\end{equation}
Then, we find a simple expression
\begin{equation}
\left(\text{Str}\mathcal{M}^2\right)_{\text{nl}}=\left[g^{a\bar a}R_{a \bar{a}S\bar S} F^{S} \bar{F}^{\bar{S}}\right]\biggl|_{S=0}.\label{glsimp}
\end{equation}

\section{Supergravity}\label{sec4}
In this section, we show the supertrace mass formula in the supergravity case. The component action of the most general supergravity system coupled to matter and a nilpotent superfield is shown in \cite{Schillo:2015ssx}. Even in the case with a nilpotent superfield, the standard supergravity formula can be applied although we have to take into account the condition $S=0$ and the additional mass for $\chi^{S}$. The scalar potential is given by
\begin{equation}
V|_{S=0}=e^K\left(D_\alpha Wg^{\alpha\bar \beta}D_{\bar\beta}\bar W-3|W|^2\right)\Biggr|_{S=0},
\end{equation}
and the mass of the spin-1/2 fermion in the Lagrangian is
\begin{equation}
\tilde{m}_{\alpha\beta}=e^{\frac K2}\left(W_{\alpha\beta}+K_{\alpha}W_{\beta}+K_{\beta}W_{\alpha}+K_{\alpha\beta}W+K_\alpha K_\beta W-\Gamma_{\alpha\beta}^\gamma D_{\gamma}W\right)\Biggr|_{S=0},\label{sgfm}
\end{equation}
where $D_\alpha W\equiv W_\alpha+K_\alpha W$ is the K\"ahler covariant derivative. The mass of the gravitino is given by
\begin{equation}
m_{3/2}=e^{\frac K2}W\biggr|_{S=0}.\label{grvtm}
\end{equation}
However, there are mass mixing terms between the gravitino $\psi_\mu$ and spin 1/2 fields $\chi^\alpha$ in the Lagrangian, $\frac1{\sqrt 2}e^{K/2}D_\alpha W \bar\psi_\mu \gamma^\mu \chi^{\alpha}$. In order to sum up the masses of fields for each spin separately, such a mixing term should be removed by diagonalizing the fermions as performed in \cite{Ferrara:2016ntj}. The procedure is the same even if there is a nilpotent superfield, and hence the mass formula for spin 1/2 components is
\begin{equation}
m_{\alpha\beta}=\tilde{m}_{\alpha\beta}-\frac{2}{m_{3/2}X}e^{K}D_\alpha WD_\beta W\Biggr|_{S=0},
\end{equation}
where $\tilde{m}_{\alpha\beta}$ is given in \eqref{sgfm} and $X=\frac{D_\alpha Wg^{\alpha\bar\beta}D_{\bar\beta}\bar W}{|W|^2}$. The gravitino mass is not modified by diagonalizing fermion masses and is given by \eqref{grvtm}. 

As in the global supersymmetry case, the fermion mass of the nilpotent superfield receives an additional mass contribution as shown in \eqref{deffm}, and hence
\begin{equation}
m'_{SS}=\left(m_{SS}+\frac{V_S}{F^S}\right)\Biggr|_{S=0},
\end{equation}
and other mass matrix elements of $m_{\a\b}$ are not corrected.
Note that the bosonic part of the F-term in supergravity is given by
\begin{equation}
F^\alpha=-e^{\frac K2}g^{\alpha\bar\beta}D_{\bar\beta}\bar W\biggr|_{S=0}.
\end{equation}

We find a useful expression for the first and second derivatives of the scalar potential as~\cite{Ferrara:2016ntj}
\begin{equation}
V_{\alpha} = -m_{\a\b} F^{\b},\label{id1} \\
\end{equation}
and 
\begin{align}
V_{\a \bar{\a}}=&(V+|m_{3/2}|^2)g_{\a\bar\a}+\tilde{m}_{\a\b}g^{\b\bar\b}\bar{\tilde m}_{\bar\a\bar\b}+e^{K}(-D_\a WD_{\bar\a}\bar W+R_{\a\bar\a}^{\ \ \ \b\bar\b}D_\b WD_{\bar\b}\bar W),\nonumber\\ \label{id2}
=&(V+|m_{3/2}|^2)g_{\a\bar\a}+m_{\a\b}g^{\b\bar\b}\bar{m}_{\bar\a\bar\b}-\left(\frac{2e^{K/2}}{m_{3/2}X}D_\a W\bar{F}^{\bar\b}\bar{m}_{\bar\a\bar\b}+\frac{2e^{K/2}}{\bar{m}_{3/2}X}D_{\bar\a} \bar WF^{\b}m_{\a\b}\right)\nonumber\\
&+\frac{4-X}{X}e^{K}D_\a WD_{\bar\a}\bar W+R_{\a\bar\a\b\bar\b}F^{\b}\bar{F}^{\bar\b}.
\end{align}
We also find the following identity,
\begin{equation}
m'_{S\a}F^{\a}=m_{S\a}F^{\a}+\frac{V_S}{F^S}F^{S}=0.
\end{equation}
This means that there always exists a zero mode, which corresponds to Goldstino.

Using the identities~\eqref{id1} and $\eqref{id2}$, we find the following expression (see~\cite{Ferrara:2016ntj} for the detailed derivation)
\begin{align}
\left(\frac{1}{2}\text{Str} \mathcal{M}^2\right)_{\rm lin} 
&=  N( V + |m_{3/2}|^2) + e^K R^{\bar\b \b} D_{\b} WD_{\bar\b} \bar{W} + \frac{2}{X} \left(\frac{1}{\bar{W}} V_{\a} g^{\a\bar\a}D_{\bar\a} \bar{W} + \frac{1}{W} V_{\bar\a} g^{\a\bar\a}D_{\a} W \right )\nonumber\\
&=  N( V + |m_{3/2}|^2) + R_{\bar\b \b} F^\b F^{\bar\b} + \frac{2}{X} \left(\frac{1}{\bar{m}_{3/2}} F^{\a}m_{\a\b}F^{\b} + \frac{1}{m_{3/2}}\bar{F}^{\bar\a}\bar{m}_{\bar\a\bar\b}\bar{F}^{\bar\b} \right ),
\end{align}
where $R_{\a\bar\a}=g^{\b\bar\b}R_{\b\bar\b\a\bar\a}$.
This expression gives the first term of \eqref{gen}.

Thus, we can formally write the supertrace formula as  
\begin{align}\label{supertracesugra}
\left(\frac{1}{2}\text{Str} \mathcal{M}^2\right)_{\rm nl} =&\frac12\left(\text{Str}\mathcal{M}^2\right)_{\text{lin}}\biggl|_{S=0} -\left(\frac12\Delta M_0^2+\Delta M_{1/2}^2\right)\biggl|_{S=0}\nonumber \\
=&\Biggl[  N( V + |m_{3/2}|^2) + R_{ \a\bar\a}F^\a \bar{F}^{\bar\a} +\frac{2}{X} \left(\frac{1}{\bar{m}_{3/2}} F^{\a}m_{\a\b}F^{\b} + \frac{1}{m_{3/2}}\bar{F}^{\bar\a}\bar{m}_{\bar\a\bar\b}\bar{F}^{\bar\b} \right )\nonumber 
\\& - g^{a \bar{S}} m_{a \b} g^{\b \bar{S}} \frac{V_{\bar{S}}}{\bar{F}^{\bar{S}}} - g^{S \bar{a}} \frac{V_S}{F^S}  g^{S \bar \b}  \bar{m}_{\bar{a} \bar\b} - (g^{S \bar{S}} )^2 \left( m_{SS} \frac{V_{\bar{S}} }{\bar{F}^{\bar{S}} } + \frac{V_S}{F^S} \bar{m}_{\bar{S} \bar{S}} + \frac{V_S V_{ \bar{S} }}{|F^S|^2}   \right)\nonumber\\
&  - (g^{a \bar{S} } V_{a \bar{S} } + g^{S \bar{a} } V_{S \bar{a}} + g^{S \bar{S}} V_{S \bar{S}})\Biggr]\Biggr|_{S=0} .
\end{align}
The factor $\frac{V_S}{F^S}$ can be rewritten as
\begin{equation}
\frac{V_S}{F^S}=-m_{SS}-m_{Sa}\frac{F^a}{F^S},
\end{equation}
which follows from \eqref{id1}. Also, we can use \eqref{id2} to derive $V_{S\bar{\alpha}}$ and $V_{\alpha\bar S}$ explicitly. However, this manipulation complicates the expression.

Although the supertrace formula in a general scalar background is complicated, it is somewhat simplified under the Minkowski vacuum condition $V_a=V=0$. Under this condition, one finds $m_{aS}=-\frac{m_{ab}F^b}{F^S}$.
We also find
\begin{align}
g^{a \bar{S} } V_{a \bar{S} } + g^{S \bar{a}} V_{S \bar{a}} + g^{S \bar{S}} V_{S \bar{S}}=&(3-g^{S\bar{S}}g_{S\bar{S}})|m_{3/2}|^2+g^{a\bar{S}}m_{a\b}g^{\b\bar\b}\bar{m}_{\bar S\bar\b}+g^{S\bar a}m_{S\b}g^{\b\bar\b}\bar{m}_{\bar a\bar\b}\nonumber\\
&+g^{S\bar S}m_{S\b}g^{\b\bar\b}\bar{m}_{\bar S\bar\b}+\frac{2\bar{m}_{\bar S \bar S}}{3m_{3/2}}\bar{F}^{\bar S}\bar{F}^{\bar S}+\frac{2m_{SS}}{3\bar{m}_{3/2}}F^SF^S\nonumber\\
&+\frac{2\bar{m}_{\bar S \bar b}}{3m_{3/2}}\bar{F}^{\bar S}\bar{F}^{\bar b}+\frac{2m_{Sb}}{3\bar{m}_{3/2}}F^SF^b-\frac13 F^ag_{a\bar a}\bar{F}^{\bar a}\nonumber\\
&+g^{S\bar{S}}R_{S\bar S\a\bar\a}F^{\a}\bar{F}^{\bar \a}+g^{S\bar a}R_{S\bar a\a\bar\a}F^{\a}\bar{F}^{\bar \a}+g^{a\bar S}R_{a\bar S\a\bar\a}F^{\a}\bar{F}^{\bar \a},
\end{align} 
where we have used $X=3$ following from $V=0$.
Using these expressions, we finally find that the supertrace \eqref{supertracesugra} is reduced to
\begin{align}
\left(\frac{1}{2}\text{Str} \mathcal{M}^2\right)_{\rm nl} =&\Biggl[  (N-3+g^{S\bar S}g_{S\bar S}) |m_{3/2}|^2 +g^{a\bar a} R_{a\bar a \a\bar\a}F^\a \bar{F}^{\bar\a}  + \frac{2}{3} \left(\frac{1}{\bar{m}_{3/2}} F^{a}m_{ab}F^{b} + \frac{1}{m_{3/2}}\bar{F}^{\bar a}\bar{m}_{\bar a\bar b}\bar{F}^{\bar b} \right )\nonumber 
\\&+g^{a\bar{S}}m_{a\a}g^{\a\bar a}\bar{m}_{\bar S\bar a}+g^{S\bar a}m_{S a}g^{ a\bar\a}\bar{m}_{\bar a\bar\a}+g^{S\bar S}m_{S\a}g^{\a\bar a}\bar{m}_{\bar S\bar a}+g^{S\bar S}m_{Sa}g^{a\bar\a}\bar{m}_{\bar S\bar\a}\nonumber\\
& -g^{a \bar{S}}g^{\a \bar{S}} m_{a \a} \frac{\bar{m}_{\bar a\bar b}\bar{F}^{\bar a}\bar{F}^{\bar b}}{(\bar{F}^{\bar{S}})^2}-g^{S \bar{a}} g^{S \bar \b}  \bar{m}_{\bar{a} \bar\b}\frac{m_{ab}F^aF^b}{(F^S)^2}-(g^{S \bar{S}} )^2\frac{|m_{ab}F^aF^b|^2}{|F^S|^4}+\frac13 F^ag_{a\bar a}\bar{F}^{\bar a}\Biggr]\Biggr|_{S=0} .\label{sugragen}
\end{align}
The first and second terms are similar to the standard supertrace formula, and other terms appear as consequences of the decoupled scalar in $\hat{S}$ and the nontrivial mass terms of $\chi^S$. In the following, we will consider simplification of this formula by imposing additional conditions on the system. 
\subsection{Simplification: no K\"ahler mixings $g_{a\bar S}=g_{S\bar a}=0$}
As seen in the previous section, the supertrace formula becomes very complicated in the presence of a nilpotent superfield. The completely general formula is not necessary, and the simplified one under certain conditions would be practically useful. Here we consider the case with the following K\"ahler potential,
\begin{equation}
K=K_0(z^a,\bar{z}^{\bar a})+K_2(z^a,\bar{z}^{\bar a})S\bar S,\label{cond1}
\end{equation}
which implies $g_{a\bar S}=g_{S\bar a}=0$ on the $S=0$ hypersurface. This condition reduces the terms in the general formula~\eqref{supertracesugra}, and we find
\begin{align}
\left(\frac{1}{2}\text{Str} \mathcal{M}^2\right)_{\rm nl} =& \Biggl[ N( V + |m_{3/2}|^2) + R_{ \a\bar\a}F^\a \bar{F}^{\bar\a} +\frac{2}{X} \left(\frac{1}{\bar{m}_{3/2}} F^{\a}m_{\a\b}F^{\b} + \frac{1}{m_{3/2}}\bar{F}^{\bar\a}\bar{m}_{\bar\a\bar\b}\bar{F}^{\bar\b} \right ) \nonumber \\
&+(g^{S\bar S})^2\left(m_{SS}\bar{m}_{\bar S\bar S}-\frac{F^a\bar{F}^{\bar a}}{F^S\bar{F}^{\bar S}}m_{Sa}\bar{m}_{\bar S\bar a}\right)-  g^{S \bar{S}} V_{S \bar{S}}\Biggr]\Biggr|_{S=0},
\end{align}
where we have used $\frac{V_S}{F^S}=-m_{SS}-m_{Sa}\frac{F^a}{F^S}$. Also, the condition~\eqref{cond1} reads $g^{S\bar S}=(g_{S\bar S})^{-1}$, and we find
\begin{align}
g^{S\bar S}V_{S\bar S}=&V+|m_{3/2}|^2+m_{S\a}g^{\a\bar \a}m_{S\bar \a}g^{S\bar S}+\left(\frac{2}{m_{3/2}X}\bar{F}^{\bar S}\bar{F}^{\bar \b}m_{\bar S\bar\b}+{\rm h.c.}\right)\nonumber\\
&+\frac{4-X}{X}F^{S}g_{S\bar S}\bar{F}^{\bar S}+g^{S\bar S}R_{S\bar S\a\bar\a}F^{\a}\bar{F}^{\bar\a}.
\end{align}
Thus, we obtain the simplified supertrace formula
\begin{align}
\left(\frac{1}{2}\text{Str} \mathcal{M}^2\right)_{\rm nl} =&  
(N-1)( V + |m_{3/2}|^2) + g^{a\bar a}R_{a\bar a \a\bar\a}F^\a \bar{F}^{\bar\a} +\frac{2}{X} \left(\frac{1}{\bar{m}_{3/2}} F^{a}m_{a\b}F^{\b} + \frac{1}{m_{3/2}}\bar{F}^{\bar a}\bar{m}_{\bar a\bar\b}\bar{F}^{\bar\b} \right ) \nonumber \\
&+(g^{S\bar S})^2\left(-\frac{F^a\bar{F}^{\bar a}}{F^S\bar{F}^{\bar S}}m_{Sa}\bar{m}_{\bar S\bar a}\right)-m_{Sa}g^{a\bar a}m_{\bar S\bar a}g^{S\bar S}-\frac{4-X}{X}F^{S}g_{S\bar S}\bar{F}^{\bar S}.\label{red}
\end{align}
Note that we have only assumed the condition~\eqref{cond1} and the vacuum condition $V_a=V=0$ is not yet imposed. Therefore, \eqref{red} is applicable to the general scalar field background in the system satisfying~\eqref{cond1}.

Let us consider the expression under the Minkowski vacuum condition $V_a=V=0$. These conditions lead to
\begin{equation}
m_{aS}=-\frac{m_{ab}F^b}{F^S},\quad X=3.
\end{equation}
Then, we find
\begin{align}
\left(\frac{1}{2}\text{Str} \mathcal{M}^2\right)_{\rm nl}\Bigg|_{\rm vac} =&  
(N-2) |m_{3/2}|^2 + g^{a\bar a}R_{a\bar a \a\bar\a}F^\a \bar{F}^{\bar\a}+\frac{\left|F^am_{ab}F^b\right|^2}{(F^{S}g_{S\bar S}\bar{F}^{\bar S})^2}\nonumber\\
&-\frac{g^{a\bar a}F^b\bar{F}^{\bar b}}{F^{S}g_{S\bar S}\bar{F}^{\bar S}}m_{ab}\bar{m}_{\bar a \bar b}+\frac13F^ag_{a\bar a}\bar{F}^{\bar a},
\end{align}
where we have used the relation $F^{S}g_{S\bar S}\bar{F}^{\bar S}=3|m_{3/2}|^2-F^ag_{a\bar a}\bar{F}^{\bar a}$ following from $V=0$.

For further simplification, we consider a special case $F^{a}=0$, which means that the Goldstino is equivalent to $\chi^S$ and there is no kinetic mixing between matter and the Goldstino $\chi^S$ by the assumption~\eqref{cond1}. In this case, the supertrace formula significantly simplifies,
\begin{equation}
\left(\frac{1}{2}\text{Str} \mathcal{M}^2\right)_{\rm nl}\Bigg|_{\rm vac}^{F^a=0} =
(N-2) |m_{3/2}|^2 + g^{a\bar a}R_{a\bar a S\bar S}F^S \bar{F}^{\bar S},\label{simplest}
\end{equation}
where we have used $F^{S}g_{S\bar S}\bar{F}^{\bar S}=3|m_{3/2}|^2$ which follows from $V=0$. In pure de Sitter supergravity~\cite{Bergshoeff:2015tra,Hasegawa:2015bza}, where $N=0$, this expression gives
\begin{equation}
\left(\frac{1}{2}\text{Str} \mathcal{M}^2\right)_{\rm nl}\Bigg|_{\rm vac} ^{\text{pure dS}}=-2|m_{3/2}|^2,
\end{equation}
which is the correct result because only the gravitino is a massive field. Note that the formula~\eqref{simplest} seems similar to the standard supertrace formula, but there are differences. At the Minkowski vacuum $V_\alpha=V=0$, the supertrace formula in linearly realized supersymmetry is given by
\begin{equation}
\left(\frac{1}{2}\text{Str} \mathcal{M}^2\right)_{\rm lin}\Biggl|_{\rm vac} =  N |m_{3/2}|^2 + R_{\a \bar \a} F^{\a}\bar{F}^{\bar \a},\label{stdvac}
\end{equation}
for the case with $N+1$ chiral multiplets. The difference between \eqref{simplest} and \eqref{stdvac} is
\begin{equation}
\Delta=2|m_{3/2}|^2+g^{S\bar S}R_{S\bar{S}S\bar S}F^S\bar{F}^{\bar{S}}.
\end{equation}
Here, although the second term vanishes if $\hat{S}$ is nilpotent, we have formally introduced it. This difference is nothing but the mass of the sGoldstino $S$,\footnote{We can identify this quantity as the sGoldstino mass by the following reason: consider a linearly realized supersymmetric system with one chiral superfield $S$. Suppose $S$ breaks supersymmetry spontaneously. The supertrace of the system is given by $\frac{1}{2}\text{Str} \mathcal{M}^2=g^{S\bar S}R_{S\bar S S\bar S}F^{S}\bar{F}^{\bar S}$. Since the Goldstino is massless, this supertrace should be $\frac{1}{2}\text{Str} \mathcal{M}^2=g^{S\bar{S}}V_{S\bar{S}}-2|m_{3/2}|^2$, where $g^{S \bar S}V_{S\bar{S}}$ is the squared mass of the sGoldstino. Then using these equations, we find $g^{S\bar{S}} V_{S\bar{S}}=2|m_{3/2}|^2+g^{S\bar S}R_{S\bar S S\bar S}F^{S}\bar{F}^{\bar S}$.} and hence \eqref{simplest} can be identified as the supertrace formula for the decoupling limit of the sGoldstino $S$.

As shown above, if only the nilpotent superfield is responsible for supersymmetry breaking and it has no mixing with matter, we find a simple and geometric expression for the supertrace formula~\eqref{simplest}. In general, however, the supertrace formula has non-geometric terms proportional to $F^a$. This shows that the spectrum of a system with a nilpotent superfield is not just a simple decoupling limit of the sGoldstino.
\section{Summary and discussion}\label{sec5}
In this work, we have derived the supertrace mass formula for a system with a nilpotent superfield in global and local four dimensional $\mathcal{N}=1$ supersymmetry. Due to the nilpotent constraint, the scalar field disappears from the physical degrees of freedom and the non-standard fermion mass~\eqref{deffm} shows up, modifying the mass of $\chi^s$. These features change the supertrace formula significantly. We have shown the general formulae \eqref{glgen} for global supersymmetry and \eqref{sugragen} for supergravity. We find that the modified supertrace formula has extra contributions, which are absent in the standard formulae. We also find that if only the nilpotent superfield is responsible for supersymmetry breaking, i.e. the F-terms of matter vanish, the supertrace formula becomes simplified significantly as shown in \eqref{glsimp} and \eqref{simplest}. For such simplified cases, the modification of the supertrace formula can be understood as the decoupling limit of $S$ in global and in local supersymmetry, respectively.

In other words, if matter fields have non-vanishing $F$-terms, we cannot interpret the nilpotent constraint as the decoupling limit of the $S$. Such an observation is consistent with the result in~\cite{Antoniadis:2011xi}, which shows that in models with multiple supersymmetry breaking fields, the decoupling limit of $S$ field leads not to a nilpotent superfield, but a different constrained superfield. In this sense, as an effective theory of a Goldstino superfield in the presence of matter, a nilpotent superfield can be a good approximation if the $F$-terms of matter fields are negligibly small.

One of the interesting examples with a nonlinear superfield is the KKLT~\cite{Kachru:2003aw}/LVS~\cite{Balasubramanian:2005zx} model in string theory. Interestingly, in such a model, supersymmetry is broken not only by a nilpotent superfield but also by moduli fields. Therefore, we need to take into account the additional corrections to the mass, which we discussed in Sec.~\ref{sec2}.

It is also interesting to consider the structure of the K\"ahler geometry in the presence of the nilpotent superfield. As we have seen, the modified supertrace formula contains the factor $F^{a}\over F^{S}$ which is absent in the standard supergravity case. One beautiful aspects of supersymmetry is that the action is described in terms of the K\"ahler geometry. However, the action with a nilpotent superfield has various non-geometric terms proportional to $F^{a}\over F^{S}$. A new supergravity formulation studied in~\cite{Freedman:2016qnq,Freedman:2017obq} might be useful to understand the geometrical meaning of the nilpotent superfield. 

\section*{Acknowledgements}

We are grateful to R. Kallosh for suggesting this project and for useful discussion and comments on the manuscript. DM and YY are supported by the SITP and by the US National Science Foundation grant PHY-1720397.

\bibliography{Supertracerefs}
\bibliographystyle{JHEP1.bst}

\end{document}